\documentclass[showpacs,preprintnumbers,amsmath,amssymb,preprint]{revtex4}
\usepackage{graphics}
\begin{document}


\title{Fragment isospin distributions and the phase diagram of
excited nuclear systems}
\author{Ad. R. Raduta}
\affiliation{National Institute of Nuclear Physics and Engineering,\\
Bucharest, POB MG6, Romania}

\begin{abstract}
Fragment average isospin distributions are investigated within
a microcanonical multifragmentation model in different regions of
the phase diagram. The results indicate that
in the liquid phase $<N/Z>$ versus $Z$ is monotonically increasing,
in the phase coexistence region it has a rise and fall shape and
in the gas phase it is constant. 
Deviations from this behavior may manifest at low fragment multiplicity
as a consequence of mass/charge conservation.
Characterization of the "free" and "bound" phases function of
fragment charge reconfirms the neutron enrichment of the "free" phase
with respect to the "bound" one irrespectively the localization of the
multifragmentation event in the phase diagram. 
\end{abstract}
\maketitle

\section{Introduction}
In recent years isospin dependent phenomena received much consideration because
of their ability to reveal information on the asymmetry term of the
nuclear equation of state (EOS).
Relying on the fact that in heavy systems in which the neutron density exceeds the
proton density the asymmetry term is repulsive for neutrons and attractive for protons,
theoretical models of heavy ion reactions predicted
different neutron composition of the liquid and vapor phases
\cite{muller,baoanli,sil,lee,ono,ditoro}.
More important for studies on EOS, the difference in chemical composition of
the gas and liquid phases during a liquid-gas phase transition reflects the magnitude
of the asymmetry term and its density dependence.
Thus, the seminal work of Mueller and Serot \cite{muller} based on
a relativistic  mean-field model of nuclear matter with arbitrary proton fraction
anticipated that is energetically more favorable for
an unstable asymmetric nuclear matter to separate into
a neutron rich low density phase and a neutron poor high density one.
Later on, 
the isospin dependent Boltzmann-Uehling-Uhlenbeck transport model \cite{baoanli},
different mean field approaches \cite{sil,lee},
the Antisymmetrized Molecular Dynamics model \cite{ono},
the Stochastic Mean Field model \cite{ditoro}, etc.
reconfirm this isospin fractionation phenomenon in both infinite and finite systems. 
Moreover, in order to offer a more realistic description of
the dynamics of charged asymmetric nuclear matter
Ref. \cite{lee} analyzes the effect of the long-range Coulomb interaction
reaching the conclusion that its effect is to diminish the isospin fractionation.

Trying to identify this process in experimental multifragmentation data,
isoscaling techniques based on grandcanonical assumptions have been applied.
The results obtained from
reactions involving different combinations of 
$^{112}$Sn and $^{124}$Sn at 50 MeV/nucleon bombarding energy \cite{xu},
$^{112,124}$Sn+$^{58,64}$Ni central collisions at 35 MeV/nucleon \cite{geraci},
multifragmentation reactions induced by high energy protons \cite{martin}
and $^{58}$Ni,$^{58}$Fe +$^{58}$Ni,$^{58}$Fe at 30, 40 and 47 MeV/nucleon \cite{shetty}
proved the expected increase of
neutron concentration in the gas phase with respect to the liquid phase.

The aim of the present work is to investigate fragment average isospin distributions
within a microcanonical multifragmentation model which includes in
a realistic way the most important ingredients of the
nuclear multifragmentation phenomenon and whose phase diagram was
studied previously.
The advantages of such a study are obvious. 
Firstly, taking into account that decaying nuclei are small isolated systems, 
a rigorous statistical treatment requires a microcanonical framework
and not an analytically tractable grandcanonical approach. 
Secondly, with respect to dynamical models, statistical models have the advantage of
dealing with precisely defined fragments.
Thus, this study is expected to offer a complementary understanding of the problem.

The paper is organized as follows. 
Section II presents the results obtained within the
Microcanonical Multifragmentation Model (MMM) \cite{mmm}
in three distinct cases: (200, 82) with and without Coulomb interaction
and (50, 23) with Coulomb interaction.
Fragments average isospin distributions are investigated as
a function of fragment charge in each situation.
Characteristic shapes of $<N/Z>$ versus $Z$ distributions are found
in each zone of the phase diagram.
Interesting finite size effects are identified for low multiplicities. 
Section III investigates the dependence of the above distributions as
a function of source isospin. 
To verify whether the observed signals survive the sequential evaporation stage,
effects of the secondary decays are discussed in Section IV. 
In order to establish a link with dynamical models predictions,
interpretation of the gas and liquid phases with respect to the
cluster size is performed in Section V.
Conclusions are drawn in Section VI.

\section{Microcanonical Multifragmentation Model predictions}

While methods to identify phase transitions in small non-extensive systems accumulate,
more importance is given to the fact that by principle the most correct statistical approach
to be used for exploding nuclei is the microcanonical one \cite{gross,mmmc}.
In the present paper the MMM version \cite{mmm} of
the microcanonical multifragmentation model \cite{smm,mmmc,randrup} is used. 

In this model fragments are placed in a spherical container defining the freeze-out volume.
All configurations allowed by mass, charge, total energy, total momentum and
total angular momentum conservation laws
and not forbidden by geometrical constraints
(overlapping between fragments or with container's walls)
are spanned by a Metropolis Monte Carlo trajectory in the configuration space.
The key quantity of the model is the weight of each configuration which has
a non-analytically tractable form and enters the expression of any physical observable.
The break-up fragments relevant for thermodynamics may be excited highly enough to
de-excite by sequential particle emission. If not explicitly mentioned otherwise,
the present study focuses on the break-up stage of the reaction
but for the sake of completeness a brief discussion of
effects of secondary decays will be included.

Depending on whether fragments are assimilated with hard non-overlapping spheres (i) or
normal nuclear density malleable objects (ii), one may distinguish two freeze-out scenarios.
Even if for a given state of the statistically equilibrated source characterized by the mass, charge,
excitation energy and freeze-out volume the two scenarios may lead to different results,
the thermodynamics associated to the model is qualitatively the same. The (ii) freeze-out scenario
has the important advantage of allowing the system to reach high densities being thus preferable
when one aims to investigate the phase diagram.

While realistic by their microcanonical foundation, statistical multifragmentation models
may be criticized because of the too simplistic treatment of the freeze-out volume.
Indeed, it is hard to imagine that fragment
production into vacuum takes place in a fixed size spherical box, but rather in a volume
fluctuating from event to event and characterized by it average value.
Statistical models used the fixed volume hypotheses in order to diminish,
presumably without significant consequences, the dimension of the configuration space.
More recent works \cite{gulminelli_const_bp}
suggest to treat the multifragmenting nucleus in a modified microcanonical ensemble in which
the volume is allowed to fluctuate and the microcanonical weight of a configuration $W(E,V)$
is multiplied by $\exp(-\beta P V)$ (where $\beta$ is the inverse temperature and
$P$ is a pressure), the average value of the volume being determined by its Lagrange multiplier
$\beta P$.

The fact that in the case of time dependent open systems
the thermodynamical definition of volume is still an open problem is illustrated by
the different concepts presently employed.
Thus, dynamical models define the freeze-out volume by the spatial extension
of the system when fragments cease to interact with each other otherwise
than by Coulomb field, implying thus a minimum distance of 2-3 fm between them.
A somehow similar image in the sense that volume does not act as an external constraint 
corresponds to the dynamical models which define the freeze-out volume with respect to
the freeze-out time, a notion which assumes chemical equilibrium: fragments can still
exchange nucleons but their multiplicity has to be time independent.
Maybe one of the most illustrative examples on what freeze-out volume may mean
within dynamical models
in contrast to the statistical ones, is given by the recent study of Ref. \cite{parlog}.
Here authors show that freeze-out volume depends dramatically on freeze-out instant and
fragment multiplicity.
As the message of the present work relies on the thermodynamical characterization
of the nuclear system,
we stress that in the case of MMM the volume is even more than an un-physical fictitious
container which obliges the {\it pre-formed} fragments not to separate,
but dictates also their partition as volume enters
the statistical weight of a configuration. 
Apart the obvious explicit dependence,
volume acts via the thermal kinetic energy defined as the difference
between the total available energy (input quantity) and all other partial energies
(internal excitation, Coulomb interaction and fragment formation $Q$).

The phase diagram associated with MMM was studied in Ref. \cite{prl2001,prl2003} and the
conclusions must be underlined. 
For small systems, like (50, 23), irrespectively whether the Coulomb interaction is
present or not, the system evolves from the liquid phase present at low excitation energies
to the gas phase corresponding to vaporized matter by crossing the coexistence zone.
For large systems, like (200, 82), which experience stronger Coulomb fields,
the situation becomes more interesting.
When one turns the Coulomb interaction off,
the system exhibits the same behavior as a small system.
When the Coulomb field is activated
the critical temperature and pressure decrease such that,
for freeze-out volumes up to about $100V_0$,
the system may evolve from
the liquid phase to gas or supercritical fluid without crossing the
phase coexistence zone \cite{prl2003}.

In the following we shall present the MMM predictions on fragment average isospin
distributions in different points of the nuclear phase diagram and
stress the fact that $<N/Z>$ versus $Z$ manifests different
behavior in the liquid, phase coexistence and gas regions.
As a general comment, we mention that the investigation of the phase space
along constant $\beta P$ paths was arbitrary and the description of the system
within a modified microcanonical ensemble with fluctuating volume 
\cite{gulminelli_const_bp} is not essential for the conclusions of the present study.
Thus, the same behavior of $<N/Z>$ vs. $Z$ distributions determined by the
event localization inside the phase diagram was obtained for
constant volume approximation (standard microcanonical approach).  

\subsection{(200,82) without Coulomb interaction}

\begin{figure}
\resizebox{0.99\textwidth}{!}{%
  \includegraphics{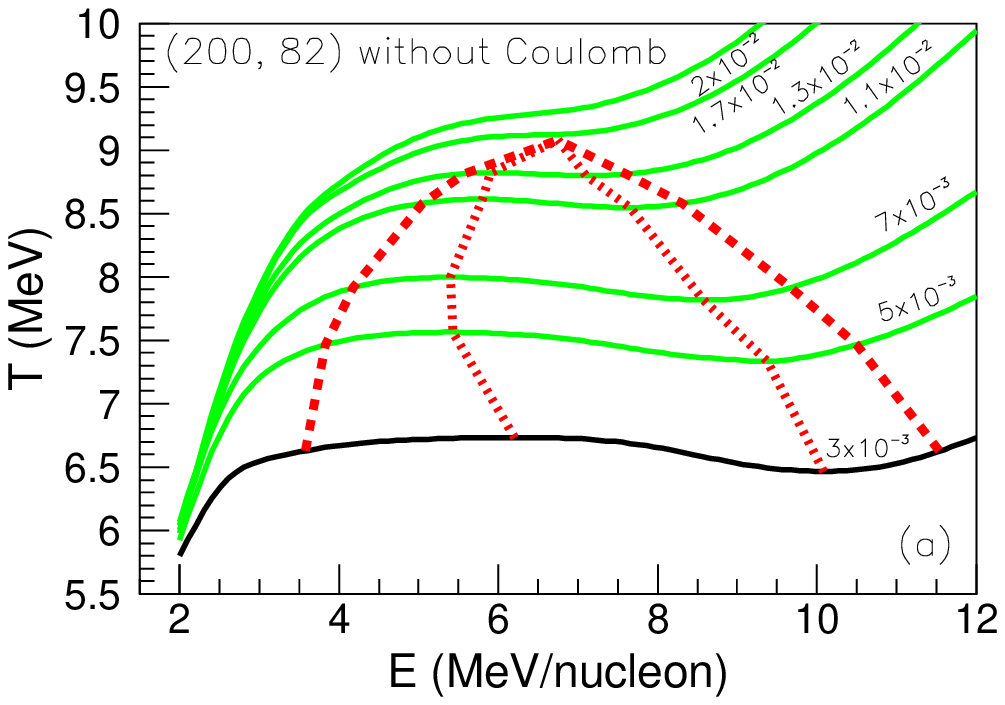}
  \includegraphics{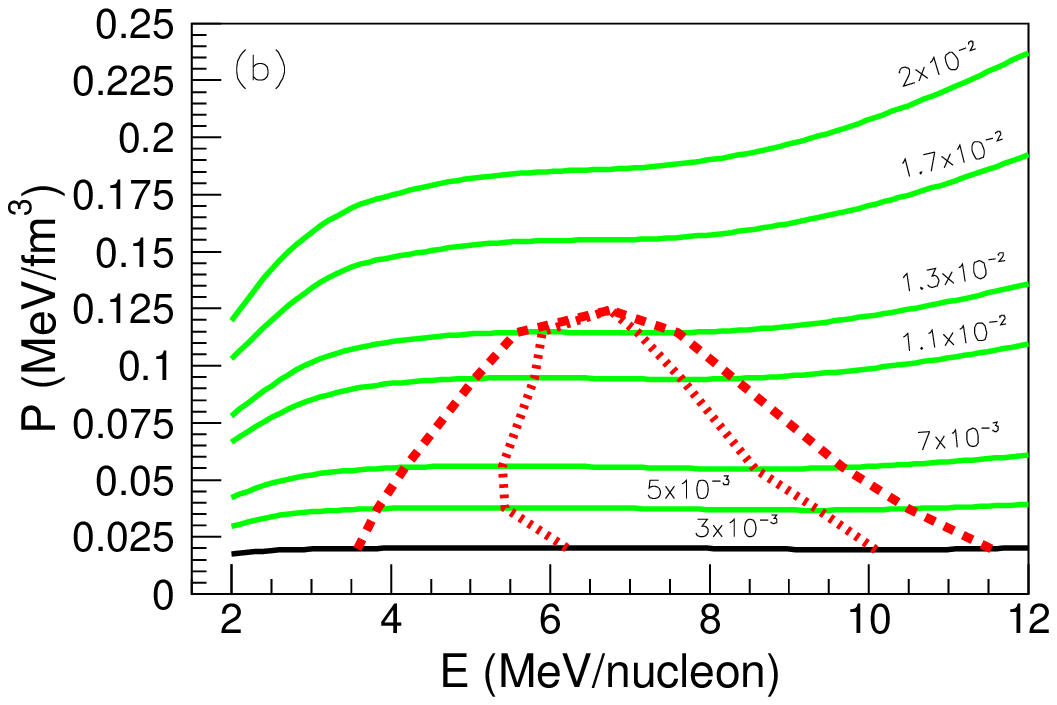}}
\resizebox{0.5\textwidth}{!}{%
  \includegraphics{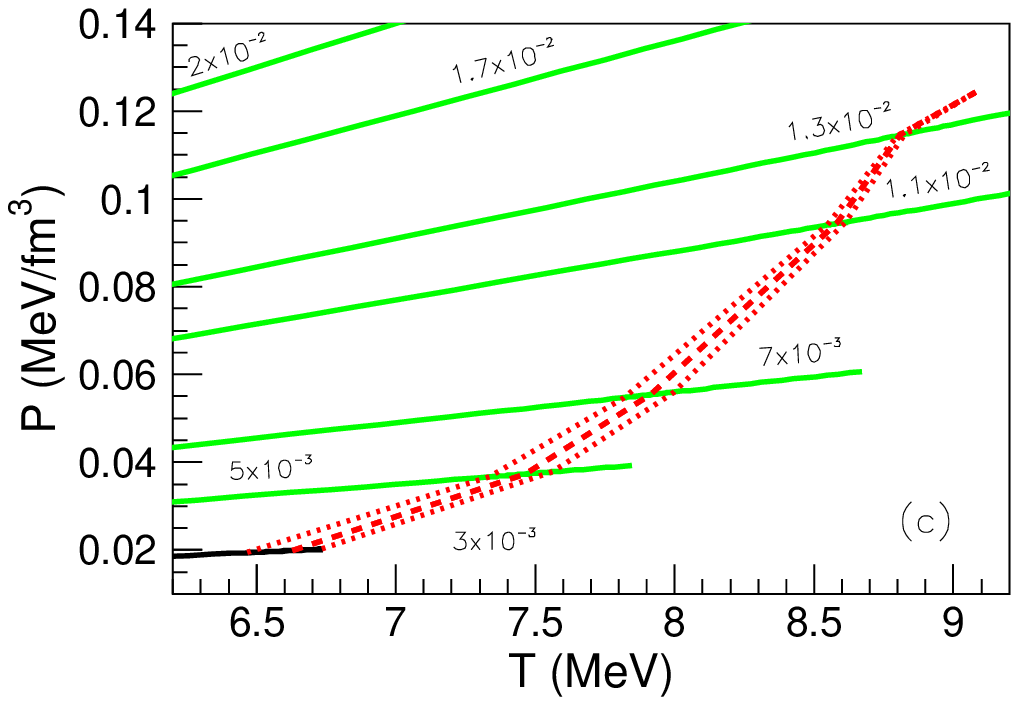}
  
  }  
\caption{Phase diagram of (200, 82) nuclear system
without Coulomb and hard-core interactions
in temperature-excitation energy (a),
pressure-excitation energy (b) and
pressure-temperature (c) representations.
The solid lines correspond to the considered iso-$\beta P$ trajectories.
The dashed lines indicate the borders of the phase coexistence region;
the dotted lines indicate the borders of the spinodal region.
The $\beta P$ values of the iso-$\beta P$ curves are labeled
in units of fm $^{-3}$.}
\label{fig:200_-c_phd}
\end{figure}

The phase diagram of the nuclear system (200, 82) without Coulomb
and hard-core interactions ((ii) freeze-out scenario)
is represented in Fig. \ref{fig:200_-c_phd}
in the temperature-excitation energy, pressure-excitation energy and
pressure-temperature planes.
The solid lines represent iso-$\beta P$ trajectories for different
values of $\beta P$ ranging from $3 \cdot 10^{-3}$ fm$^{-3}$ to
$2 \cdot 10^{-2}$ fm$^{-3}$, as indicated on the figure. 
The borders of the phase coexistence region were evaluated using
Maxwell construction on the iso-$\beta P$ caloric curves
and are plotted with dashed lines.
The borders of the spinodal region are defined
as the locus of the inflexion points of $T(E)|_{\beta P}$ curves and
are plotted with dotted lines.
The critical point  is characterized by the following set of values:
$T_C$=9.1 MeV, $P_C=1.25 \cdot 10^{-1}$ MeV/fm$^3$, $E_C$=6.75 MeV/nucleon
and $(V/V_0)_C$=1.33.
 
The shapes of $<N/Z>$ versus $Z$ distributions have been investigated
along all iso-$\beta P$ paths represented in Fig. \ref{fig:200_-c_phd}.
To illustrate the conclusions, we scan the phase space along the trajectory
characterized by $\beta P=3 \cdot 10^{-3}$ fm$^{-3}$. The states accessed in this way
are similar to the ones obtained in nuclear multifragmentation reactions.
Thus, as the excitation energy increases
from 2 to 12 MeV/nucleon, the temperature ranges around 6 MeV and the
average freeze-out volume increases linearly from 2$V_0$ to 16$V_0$. 

\begin{figure}
\resizebox{0.55\textwidth}{!}{%
  \includegraphics{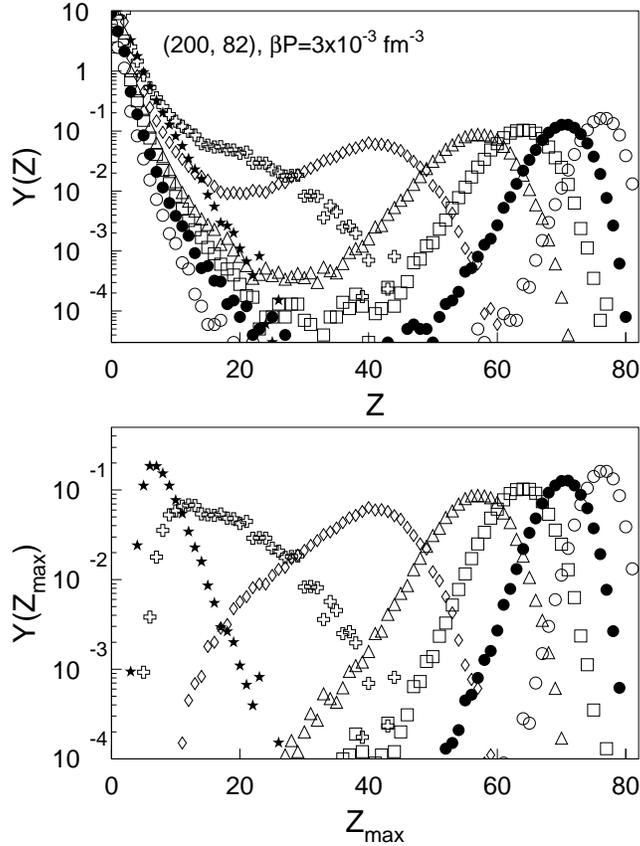}}
\caption{Fragment charge distributions (upper panel) and 
charge distributions of the largest fragment (lower panel)
for the (200, 82) nuclear system without Coulomb interaction
at different excitation energies.
For all considered situations the system in constrained by
$\beta P=3 \cdot 10^{-3}$ fm$^{-3}$.
The key legend is the following: 
open circles $E$=3 MeV/nucleon,
filled circles $E$=4 MeV/nucleon,
open squares $E$=5 MeV/nucleon,
open triangles $E$=6 MeV/nucleon,
open diamonds $E$=8 MeV/nucleon,
crosses $E$=10 MeV/nucleon,
stars $E$=12 MeV/nucleon.
}
\label{fig:200_-c_z}
\end{figure}

Once clarified the thermodynamical behavior of the system and before investigating
the fragment average isospin distributions as a function of fragment size,
it is useful to have
a clear picture on the fragment size distributions produced in the considered
multifragmentation events. Fig. \ref{fig:200_-c_z} depicts 
the fragment charge distributions (upper panel) and charge distributions of the largest
fragment in each event (lower panel).

Even if, our main purpose for plotting $Y(Z)$ is only to
illustrate fragment charge population for each state of the source, the correct information
on event localization in the phase diagram one may get in this case 
from the shape of $Y(Z)$ distributions is remarkable.
Thus, as one may notice from the upper panel of Fig. \ref{fig:200_-c_z},
as far as the system consists of liquid + under-saturated vapor ($E$=3.6 - 11.5 MeV/nucleon)
$Y(Z)$ has a U- or a shoulder-like shape
while it falls exponentially in the super-saturated vapor phase ($E>$ 11.5 MeV/nucleon), 
as anticipated in the early days of multifragmentation \cite{smm,mmmc,randrup}.
This result is striking the more so as it does not hold for the case in which the
Coulomb interaction is strong, as one may see in the next subsection.
 
The decision to investigate how $Y(Z_{max})$ distributions look like is motivated by the fact
that the largest fragment is expected to be an order parameter of the phase
transition \cite{campi} and a good estimation of the liquid phase 
\cite{purdue}. 
The information emerging from the lower panel of
Fig. \ref{fig:200_-c_z} is that by increasing the excitation energy 
the centroid of the $Y(Z_{max})$ distribution shifts toward lower values while its
shape evolves in a complicated manner.
Thus, despite in the coexistence region, for $E$=4, 5 and 6 MeV/nucleon
$Y(Z_{max})$ is single peaked and symmetric enough to hinder any 
information on the true localization of the event in the phase diagram of the system.
For higher excitation energies, $E$=8, 10 MeV/nucleon, the $Y(Z_{max})$ distributions become
broad and asymmetric such that one may interpret their shapes as the result of the superposition
of two Gaussian distributions corresponding to each phase of the system.
For this short energy interval the information extracted from
$Y(Z_{max})$ distributions is correct.
For excitation energies higher than 11.5 MeV/nucleon
the system consists of super-saturated vapor, 
the $Y(Z)$ distribution has an exponential decrease, the largest
residual nucleus has on average less than 10 protons and the
$Y(Z_{max})$ distribution is single peaked and narrow.
For this case the information inferred from $Y(Z_{max})$ is again correct.
The limited and not always correct information the largest cluster distribution
may provide on localization of multifragmentation events in the phase diagram
has been recently studied within a Lattice Gas Model \cite{gulminelli} and explained as
a consequence of charge/mass conservation in microcanonical and canonical ensembles.

\begin{figure}
\resizebox{0.6\textwidth}{!}{%
  \includegraphics{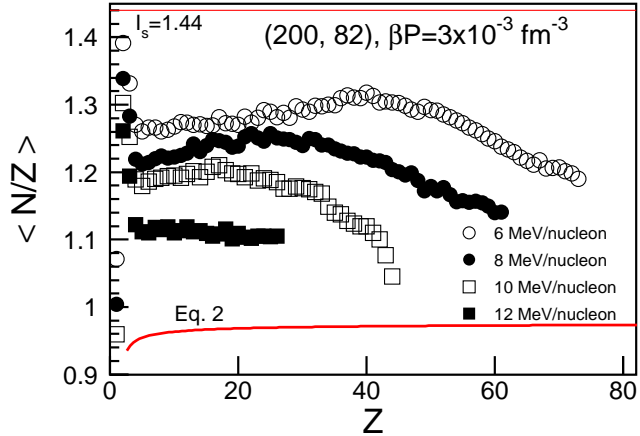}}
\caption{Fragment average isospin distributions as a function of
fragment charge for the (200, 82) nuclear system without Coulomb interaction
at different excitation energies.
For all considered situations the system in constrained by
$\beta P=3 \cdot 10^{-3}$ fm$^{-3}$.
The horizontal line indicates the source's isospin, $I_s$=1.44.
The thick solid line corresponds to the most probable isospin value obtained
from the liquid-drop formula of the binding energy, Eq. \ref{eq:i_a}.}
\label{fig:200_-c_i}
\end{figure}

Fig. \ref{fig:200_-c_i} represents the fragment average isospin ($<N/Z>$) distribution
as a function of fragment charge
for the above considered excitation energies. Letting apart the 
average isospin of light charged particles ($Z<5$) strongly affected by structure effects,
one may observe that in the coexistence region $<N/Z>$ vs. $Z$ manifests a clear
rise and fall behavior and in the super-saturated vapor phase it is constant. 
Moreover, in the coexistence region the average isospin of a fragment belonging to the
liquid is monotonically decreasing with its charge. 

The occurrence of the liquid + under-saturated vapor
phase at low values of excitation energy and/or freeze-out volume
prevents fragment production in the intermediate size domain ($15<Z<30$), as evidenced
in Fig. \ref{fig:200_-c_z} by the $Y(Z)$ distribution corresponding to $E$=3 MeV/nucleon.
Moreover, the small values of the total fragment multiplicity make the isospin
of clusters whose size is close to the source size sensitive to mass/charge conservation.
For these reasons, it was not possible to access conclusive $<N/Z>$ vs. $Z$ distributions.

The effect of excitation energy on $<N/Z>$ vs. $Z$ distributions within a given region of the
phase diagram is quite trivial. Thus, by increasing the excitation energy the number of
emitted neutrons increases such that for conserving the mass and charge of the
total system the isospin of the rest of matter is decreasing leading to the
observed shift of $<N/Z>$ vs. $Z$ distributions toward lower values.
The decrease of the turning point where the rising distribution starts to
decrease may be understood having in mind the diminish of the
liquid part with the excitation energy increase.

\subsection{(50,23) with Coulomb interaction}

To verify whether the obtained results stand valid while modifying the system size
and switching the Coulomb interaction on,
in the following we shall investigate another system,
(50, 23) with Coulomb interaction.

\begin{figure}
\resizebox{0.99\textwidth}{!}{%
  \includegraphics{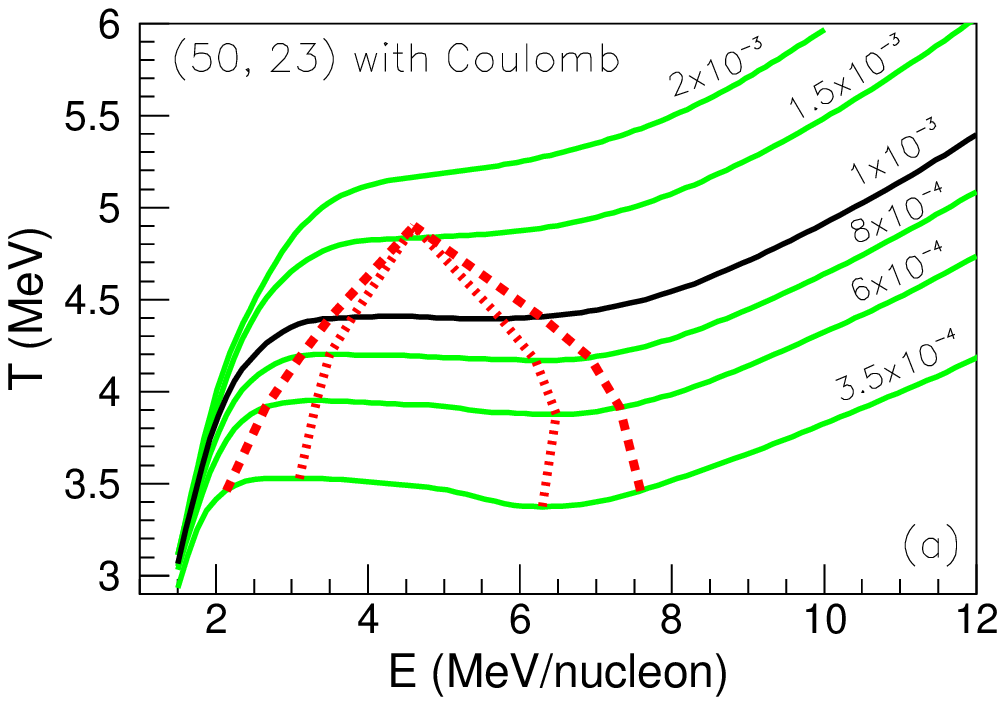}
  \includegraphics{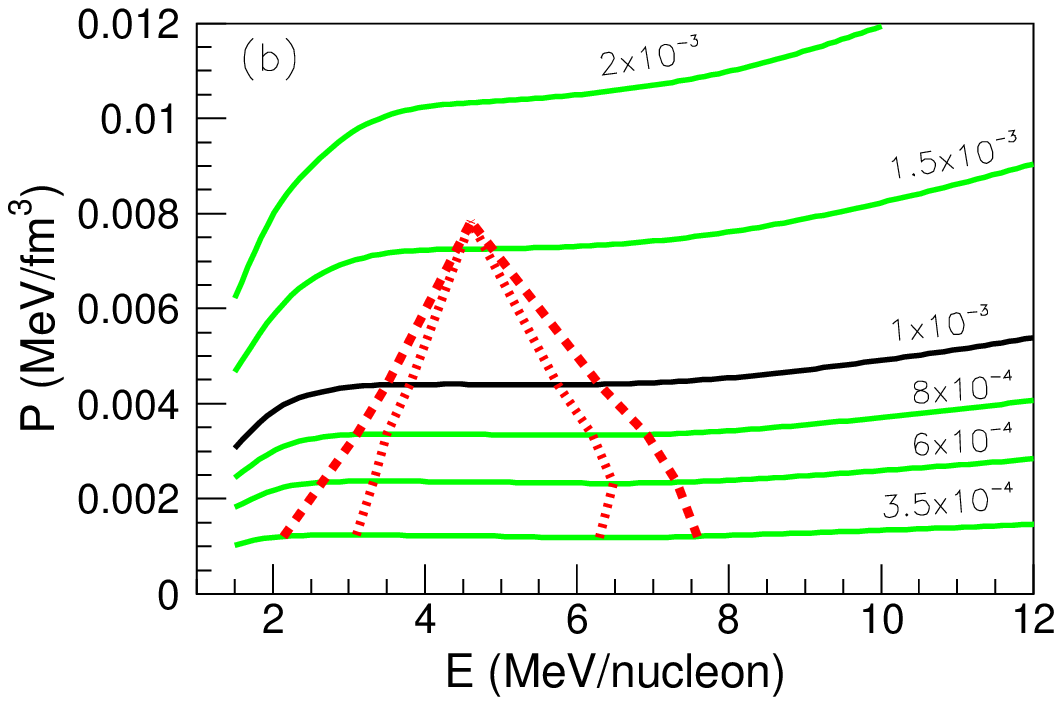}}
\resizebox{0.49\textwidth}{!}{%
  \includegraphics{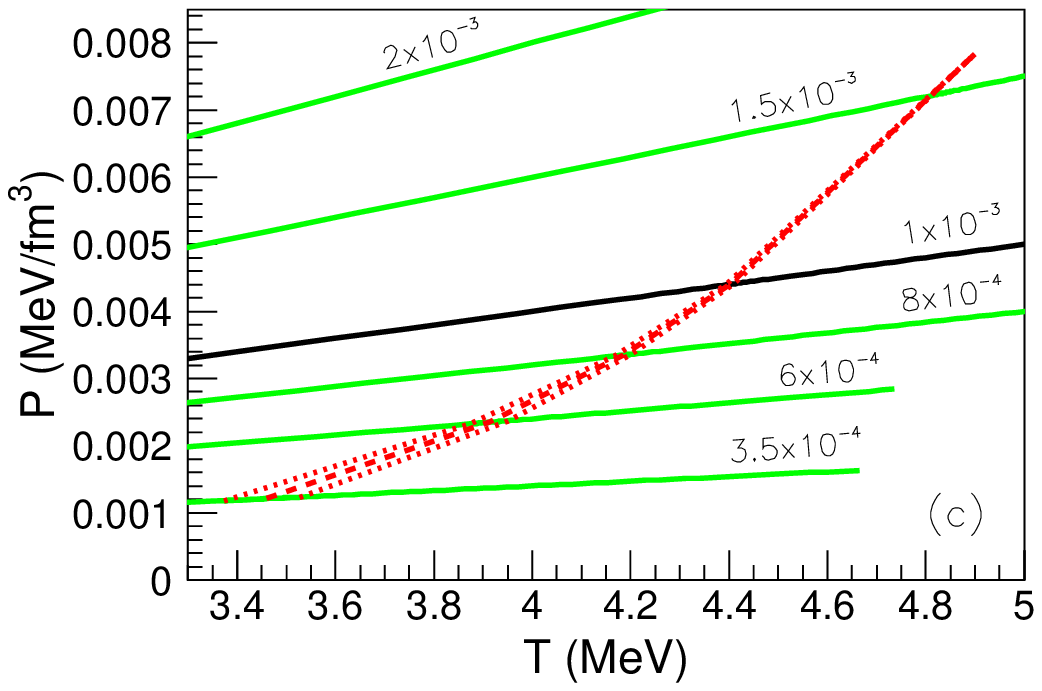}}    
\caption{Phase diagram of (50, 23) nuclear system with Coulomb interaction and
free-volume parameterization in temperature-excitation energy (a),
pressure-excitation energy (b) and
pressure-temperature (c) representations.
The solid lines correspond to the considered iso-$\beta P$ trajectories.
The dashed lines indicate the borders of the phase coexistence region;
the dotted lines indicate the borders of the spinodal region.
The $\beta P$ values of the iso-$\beta P$ curves are labeled
in units of fm $^{-3}$.}
\label{fig:50_phd}
\end{figure}

The phase diagram obtained using the (ii) freeze-out scenario
is represented in Fig. \ref{fig:50_phd} 
in the temperature-excitation energy,
pressure-excitation energy and pressure-temperature planes.
As in the previous case, the solid lines represent
iso-$\beta P$ trajectories for different values of $\beta P$ ranging from
$3.5 \cdot 10^{-4}$ fm$^{-3}$ to $2 \cdot 10^{-3}$ fm$^{-3}$,
as indicated on the figure. 
The borders of the phase coexistence (dashed lines) and
the spinodal (dotted lines) regions are determined as in subsection A. 
The critical point is characterized by the following
set of values:
$T_C$=4.9 MeV, $P_C=7.8 \cdot 10^{-3}$ MeV/fm$^3$,
$E_C$=4.6 MeV/nucleon and $(V/V_0)_C$=12.
The striking difference between the coordinates of this critical point and the one
corresponding to the (200,82) without Coulomb case is not only a finite-size effect,
but mainly the consequence of including the Coulomb interaction.
Thus, even for a relatively small system like the (50, 23) nucleus,
the Coulomb field is strong enough to force the system to occupy a large volume and to diminish
in this way the repulsive effect.
If true, a critical volume of about 12$V_0$ would imply that real nuclear multifragmentation
reactions for which the freeze-out volume was estimated to be 3-9 $V_0$ take place
at super-critical values. 
 
As in the previous case, to illustrate the behavior of
fragment average isospin distributions in the liquid, phase coexistence and gas regions
we follow a constant $\beta P=1 \cdot 10^{-3}$ fm$^{-3}$ path.
As one may see in Fig. \ref{fig:50_phd}, along the considered trajectory
the system consists of a liquid with an under-saturated vapor for $E \leq$ 3.5 MeV/nucleon;
for 3.5 MeV/nucleon $\leq E \leq$ 6.3 MeV/nucleon the system undergoes phase
separation into a liquid and its associated saturated vapor; and for
$E >$ 6.3 MeV/nucleon the system is a super-saturated vapor. 
To have complete thermodynamical information on the explored states we mention that the
excitation energy increase from 1.5 to 8 MeV/nucleon leads to a linear increase of
the average freeze-out volume from 10$V_0$ to 50$V_0$.

\begin{figure}
\resizebox{0.55\textwidth}{!}{%
  \includegraphics{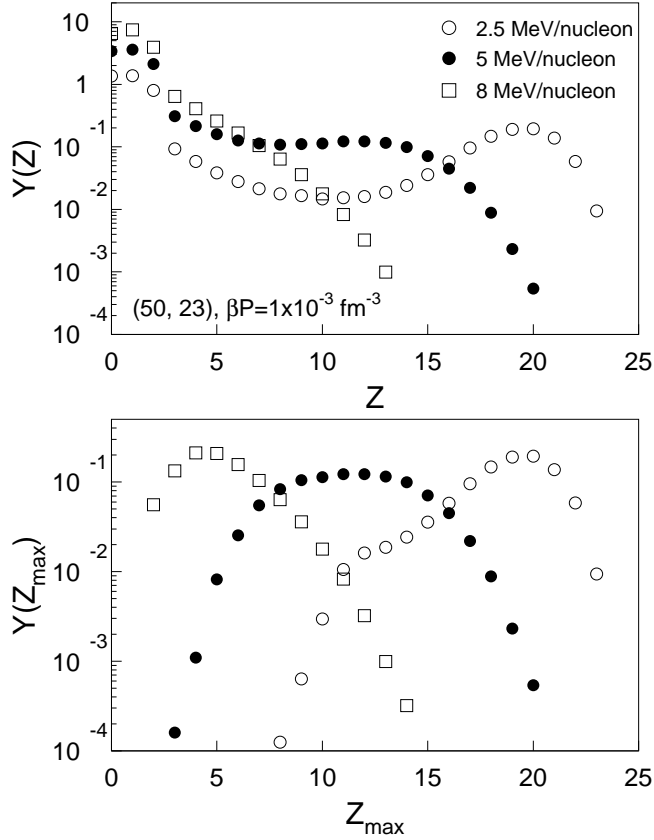}}
\caption{Fragment charge distributions (upper panel) and 
charge distributions of the largest fragment (lower panel)
for the (50, 23) nuclear system with Coulomb interaction
at different excitation energies.
For all considered situations the system in constrained by
$\beta P=1 \cdot 10^{-3}$ fm$^{-3}$.}
\label{fig:50_z}
\end{figure}

Fragment charge distributions and charge distributions of the
largest fragment are plotted in Fig. \ref{fig:50_z}.
As in the previous case, the localization of multifragmentation
events in the phase space is correctly indicated by the shape of
the $Y(Z)$ distributions: up to 3.5 MeV/nucleon $Y(Z)$ has a
U-shape, for 3.5 $<E<$ 6 MeV/nucleon $Y(Z)$ has shoulder-like shape
and for $E>$ 6.5 MeV/nucleon is falling exponentially.
In what regards the $Y(Z_{max})$ distributions, the small size of the system
makes the mass/charge conservation effects stronger than those observed
for the (200, 82) nucleus. Indeed,
the $Y(Z_{max})$ distributions corresponding to 2.5 and 8 MeV/nucleon are
truncated and no bimodality nor any particular structure which may suggest
superposition of close distributions corresponding to different phases can be
identified for 5 MeV/nucleon excitation energy. 
The only effect which may suggest in this last case a particular state of the system 
is the rather broad $Y(Z_{max})$ distribution. In addition, for $E$=2.5 MeV/nucleon
the competition between evaporation-like decay and multifragmentation induces
a hump of the $Y(Z_{max})$ distribution which could be interpreted as bimodality
and erroneously associated to phase coexistence.

\begin{figure}
\resizebox{0.6\textwidth}{!}{%
  \includegraphics{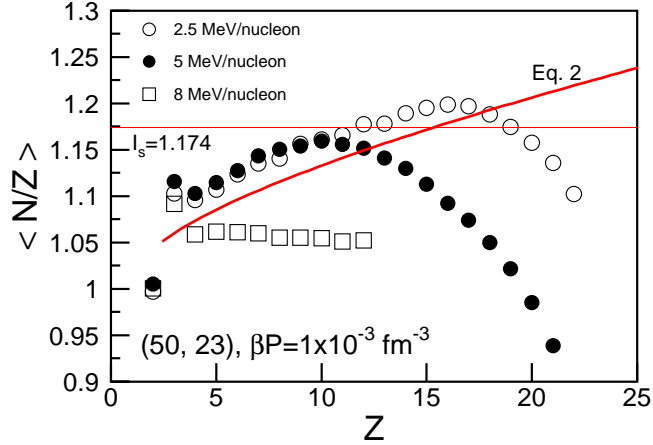}}
\caption{Fragment average isospin distributions as a function of
fragment charge for the (50, 23) nuclear system with Coulomb interaction
in the liquid ($E$=2.5 MeV/nucleon),
liquid-gas coexistence ($E$=5 MeV/nucleon)
and gas phase ($E$=8 MeV/nucleon).
For all considered situations the system in constrained by
$\beta P=1 \cdot 10^{-3}$ fm$^{-3}$.
The horizontal line indicates the source's isospin, $I_s$=1.174.
The thick solid line corresponds to the most probable isospin value obtained
from the liquid-drop formula of the binding energy, Eq. \ref{eq:i_a}.}
\label{fig:50_i}
\end{figure}

Fig. \ref{fig:50_i} presents $<N/Z>$ versus $Z$
distributions corresponding to the liquid, liquid-gas and gas phases
of the (50, 23) system. 
For the coexistence region ($E$=5 MeV/nucleon) and for the gas phase
($E$=8 MeV/nucleon) the shapes look like the ones obtained for the (200, 82) nucleus:
in the phase coexistence $<N/Z>$ versus $Z$ has a rise and fall shape and
for the gas phase it is almost constant.
A striking result is the one corresponding to 2.5 MeV/nucleon which seems to be compatible
with the phase coexistence region. 
The explanation of this apparent paradox lies in the fact that for such a small system and low
excitation energies the total fragment multiplicity is around 3, meaning that each
emitted light particle will modify drastically the isospin of the residual nucleus.
Thus, the observed rise and fall is a consequence of mass/charge conservation.

\subsection{(200,82) with Coulomb interaction}

\begin{figure}
\resizebox{0.99\textwidth}{!}{%
  \includegraphics{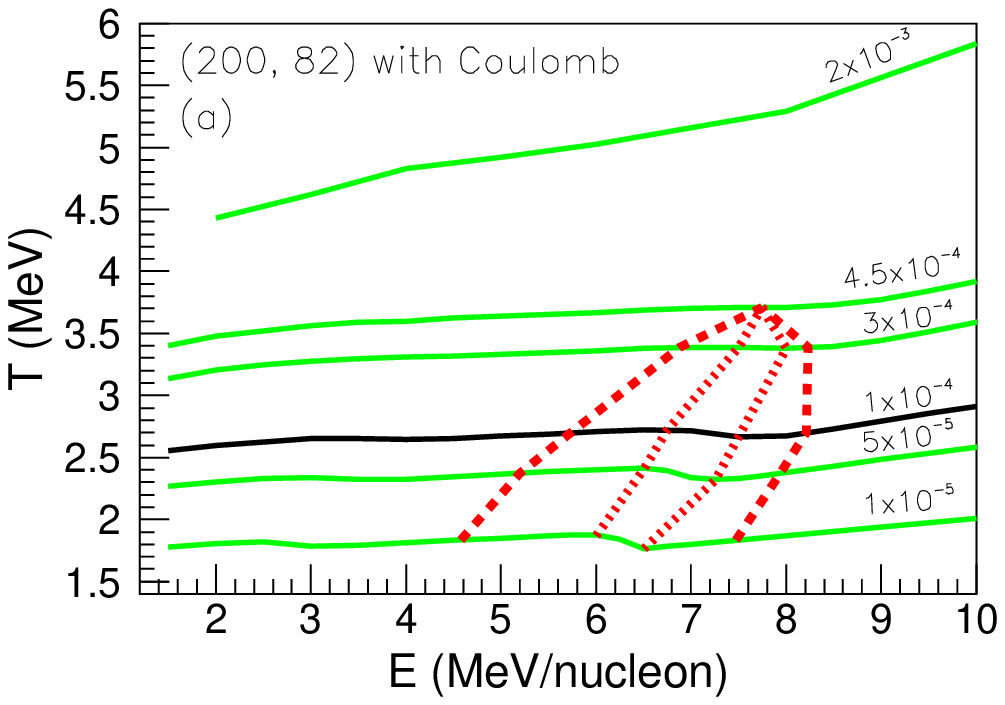}
  \includegraphics{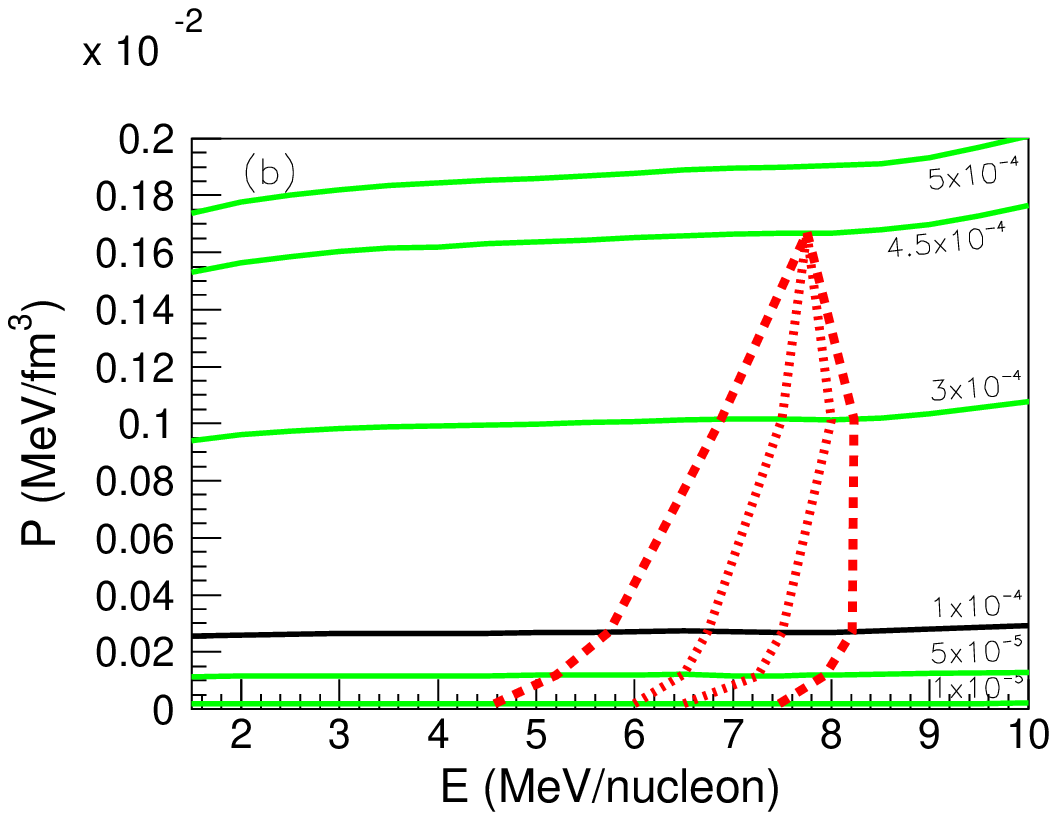}}
\resizebox{0.49\textwidth}{!}{%
  \includegraphics{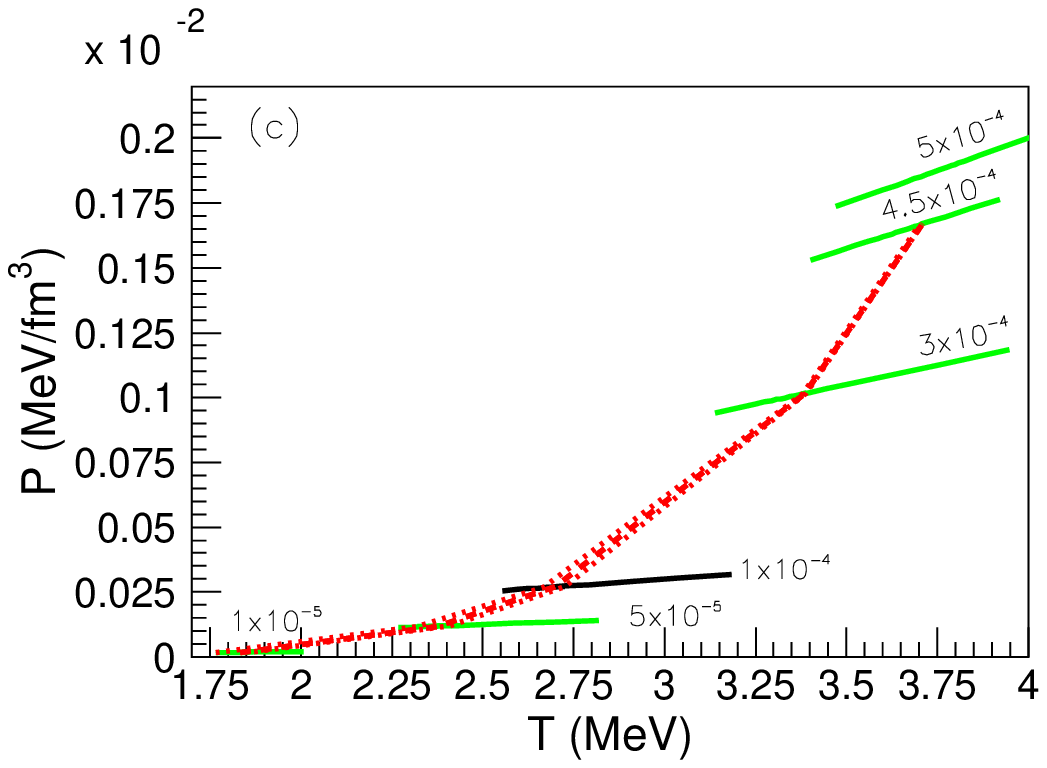}}    
\caption{Phase diagram of (200, 82) nuclear system with Coulomb interaction and
free-volume parameterization in temperature-excitation energy (a),
pressure-excitation energy (b) and
pressure-temperature (c) representations.
The solid lines correspond to the considered iso-$\beta P$ trajectories.
The dashed lines indicate the borders of the phase coexistence region;
the dotted lines indicate the borders of the spinodal region.
The $\beta P$ values of the iso-$\beta P$ curves are labeled
in units of fm $^{-3}$.}
\label{fig:200+C_phd}
\end{figure}

The system's expansion into volumes much larger than those estimated for
multifragmentation reactions under the repulsive effect of the Coulomb
field is expected to increase with the system size. This fact is confirmed
by the phase diagram of the (200, 82) excited nucleus plotted in
Fig. \ref{fig:200+C_phd} in temperature-excitation energy,
pressure-excitation energy and pressure-temperature representations.
The critical point is characterized by
$T_C$=3.7 MeV, $P_C=1.7 \cdot 10^{-3}$ MeV/fm$^3$,
$E_C$=7.75 MeV/nucleon and $(V/V_0)_C$=130 and phase coexistence occurs
for freeze-out volumes of at least few hundreds $V_0$.

\begin{figure}
\resizebox{0.99\textwidth}{!}{%
 \includegraphics{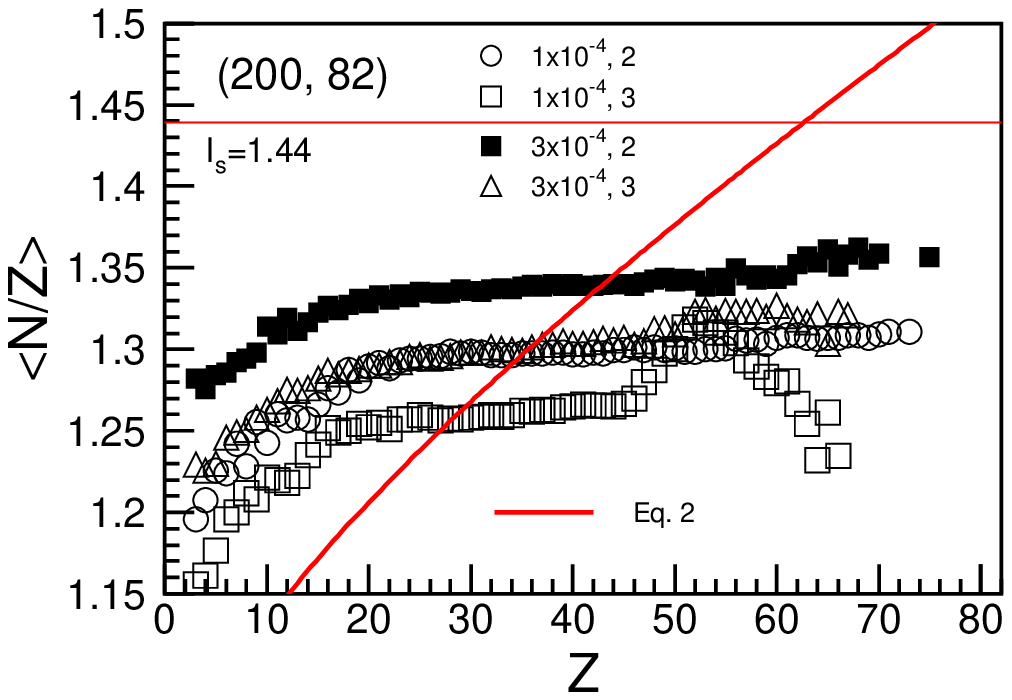}
 \includegraphics{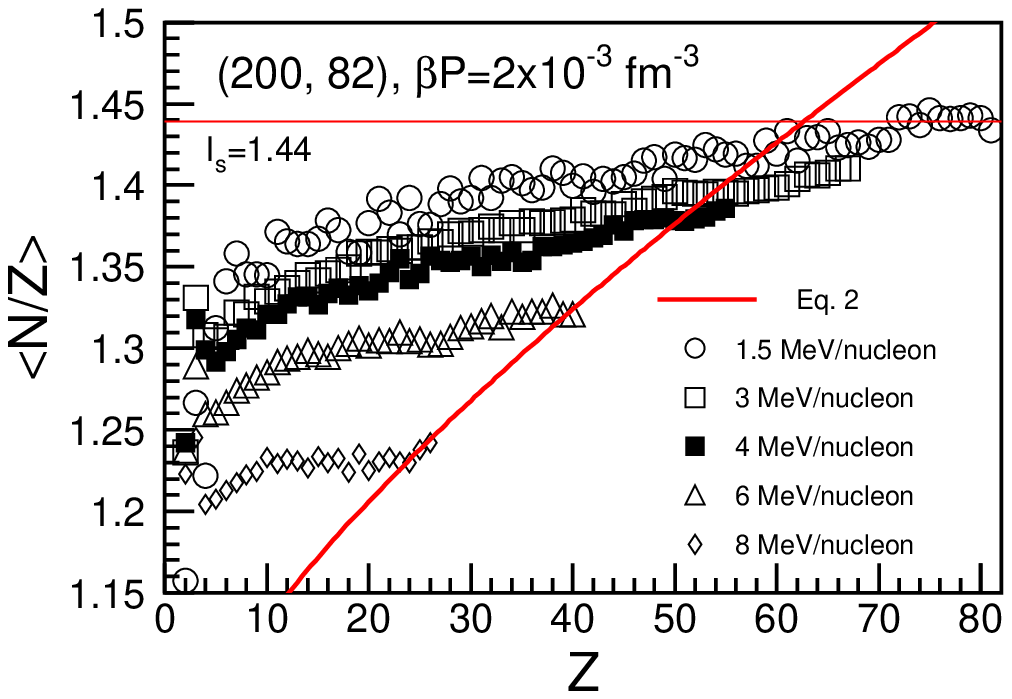}
}
\resizebox{0.5\textwidth}{!}{%
 \includegraphics{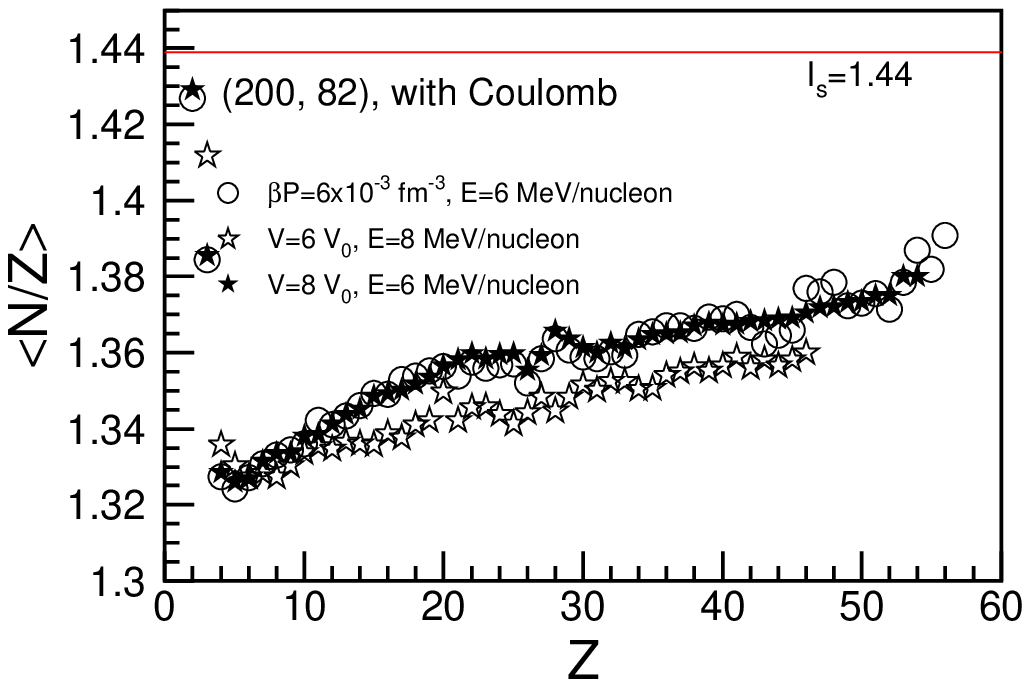}
 }
\caption{Fragment average isospin distributions as a function of
fragment charge for the (200, 82) nuclear system with Coulomb interaction.
Left-upper panel:
liquid + under-saturated vapor phase
for $\beta P=1 \cdot 10^{-4}$ and $3 \cdot 10^{-4}$ fm$^{-3}$, 
$E$=2, 3 MeV/nucleon;
right upper panel: super-critical fluid phase along
$\beta P=2 \cdot 10^{-3}$ fm$^{-3}$;
lower panel: 
(i) freeze-out scenario and different freeze-out volume approximations
as indicated on the figure.
The horizontal line indicates the source's isospin, $I_s$=1.44.
The thick line corresponds to predictions of Eq. \ref{eq:i_a}.}
\label{fig:200_+c_i}
\end{figure}

Even if not interesting for nuclear multifragmentation, the iso-$\beta P$
trajectories plotted in Fig. \ref{fig:200+C_phd} offer the possibility to investigate
the liquid phase without the undesired effects of low fragment multiplicity.
The left-upper panel of Fig. \ref{fig:200_+c_i} depicts the typical 
monotonic increase of $<N/Z>$ vs. $Z$ distributions in different states of
the liquid + under-saturated vapor system ($\beta P=1 \cdot 10^{-4}$ fm$^{-3}$,
$E$ = 2 MeV/nucleon and $\beta P=3 \cdot 10^{-4}$ fm$^{-3}$,
$E$ = 2, 3 MeV/nucleon).
Deviations from this shape may be due to mass/charge conservation, 
as one may see for $\beta P=1 \cdot 10^{-4}$ fm$^{-3}$ and $E$ = 3 MeV/nucleon  
case where the hump present around $Z$=50 is produced by the dominant
fission-like decay.
Thermodynamical states more similar to those produced in
multifragmentation reactions are accessed along the
super-critical  $\beta P=2 \cdot 10^{-3}$ fm$^{-3}$
path where the freeze-out volume increases linearly from
4$V_0$ (for $E$=1.5 MeV/nucleon) to 30$V_0$ (for $E$=8 MeV/nucleon).
The corresponding average isospin
distributions plotted in the right-upper panel of Fig. \ref{fig:200_+c_i}
show the same monotonic increase observed for the liquid phase.
It is worthwhile to mention here that a $<N/Z>$ increasing with $Z$
was reported by other microcanonical models \cite{shetty_botvina} and
experimental data obtained in deep inelastic collisions of
Kr and Nb \cite{texas}.

The lower panel of  Fig. \ref{fig:200_+c_i} presents some extra
fragment average isospin distributions corresponding to the same
(200, 82) with Coulomb system under
the (i) freeze-out scenario
and different freeze-out volume approximations:
(1) constant $\beta P=6 \cdot 10^{-3}$ fm$^{-3}$ and $E$=6 MeV/nucleon
($<V>$=8.08 $V_0$, $T$=6.88 MeV),
(2) constant $V=6 V_0$ and $E$=8 MeV/nucleon ($T$=7.68 MeV)
and (3) constant $V=8 V_0$ and $E$=6 MeV/nucleon ($T$=6.90 MeV).
The first and the third cases were chosed such that to be
characterized by almost identical values of the (average) freeze-out volume,
excitation energy and temperature but
differ by the freeze-out volume approximation: in the first case, the volume is
fluctuating while in the last one it is fixed. 
The monotonic increase of $<N/Z>$ versus $Z$ distributions obtained in all
cases and the almost perfect superposition between the isospin distributions
obtained in (1) and (3) illustrate that the presented results do not depend on
freeze-out volume approximation.

Regarding the evolution of $<N/Z>$ vs. $Z$ with increasing excitation energy,
the comments done in the previous section stand valid,
namely that by increasing the number of emitted neutrons,
the isospin of the rest of the system decreases leading to the observed shift of 
$<N/Z>$ versus $Z$ distributions.

\begin{figure}
\resizebox{0.55\textwidth}{!}{%
  \includegraphics{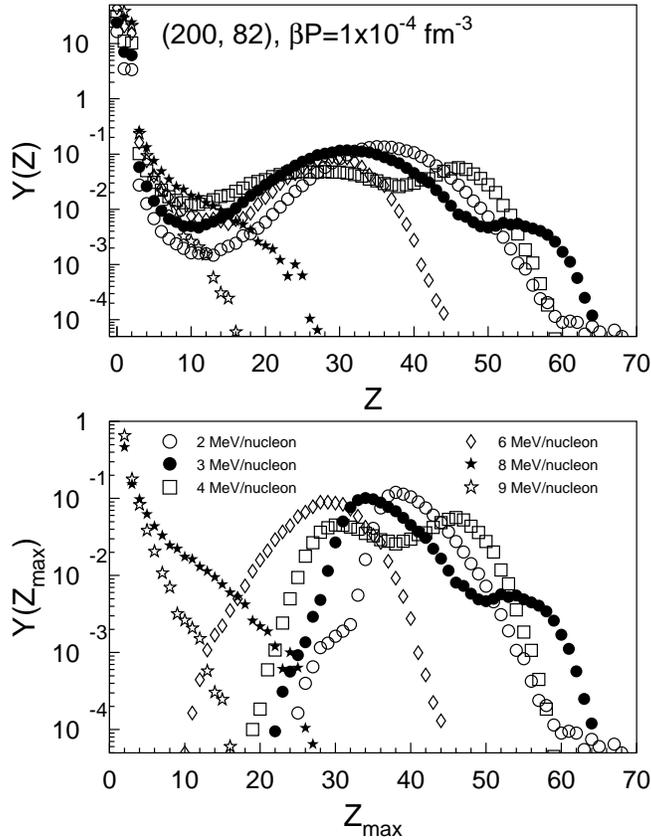}}
\caption{Fragment charge distributions (upper panel) and 
charge distributions of the largest fragment (lower panel)
for the (200, 82) nuclear system with Coulomb interaction
at different excitation energies.
For all considered situations the system in constrained by
$\beta P=1 \cdot 10^{-4}$ fm$^{-3}$.}
\label{fig:200_+c_z}
\end{figure}

New examples on how misleading fragment charge distribution may be when dealing
with microcanonical approaches are provided by the same
$\beta P$=$1 \cdot 10^{-4}$ fm$^{-3}$ path and
are plotted in Fig. \ref{fig:200_+c_z}.
Thus, for the lowest considered excitation energy, 2 MeV/nucleon,
the system consists of liquid and under-saturated vapor, while the shoulder-like
shape of $Y(Z)$ distribution and the asymmetric shape of $Y(Z_{max})$ distribution
would suggest liquid-saturated vapor coexistence.
For 3 and 4 MeV/nucleon the situation is even more difficult: the presence of
fission as decay mechanism induces a W-shape of the $Y(Z)$ distributions
and a bimodal structure of $Y(Z_{max})$ distributions which could be erroneously
interpreted as a signature of phase coexistence. When the system enters the
liquid-saturated vapor state, the situation is again tricky: for 6 MeV/nucleon the
$Y(Z_{max})$ distribution is single-peaked. The high energy border of the phase
coexistence region (8 MeV/nucleon) is characterized by a shoulder-like shape of
both $Y(Z)$ and $Y(Z_{max})$ distributions.
In the supersaturated vapor phase, $Y(Z)$ and $Y(Z_{max})$ fall exponentially,
giving for the first time correct information on the localization of the multifragmentation
event in the phase space.

To understand whether the obtained behavior of the fragment average
isospin distributions is caused by the specific fragment partition in
different regions of phase space or a trivial consequence of the
liquid-drop binding energy,
\begin{equation}
B(A,Z)=a_v \left( 1-a_i \left (1-\frac{2Z}{A} \right )^2\right) A-
a_s \left( 1-a_i \left(1-\frac{2Z}{A} \right)^2\right) A^{2/3}-
a_c Z^2 A^{-1/3}+a_a Z^2/A, 
\end{equation}
one may calculate the most probable value of a fragment isospin
as a function of its mass (charge) by requiring that
$\partial B/\partial I|_A=0$.
In the present simulation, $a_v$=15.4941 MeV, $a_s$=17.9439 MeV,
$a_i$=1.7826, $a_c$=0.7053 MeV and $a_a$=1.1530 MeV \cite{ld}.

Solving this equation one obtains for the most probable isospin,
\begin{equation}
I(A)=1+\frac{a_c A^{5/3}-a_a A}{2 a_i \cdot (a_v A-a_s A^{2/3})},
\label{eq:i_a}
\end{equation}
a function monotonically increasing with fragment size, irrespectively
whether $a_c$=0 (the Coulomb interaction is switched off) or not.
The corresponding isospin distributions as a function of $Z$
obtained using Eq. \ref{eq:i_a} are plotted in Figs. 3, 6 and 8.
Thus, the only situation where we obtain qualitative agreement between
the MMM fragment average isospin distributions and predictions of the liquid-drop
binding energy corresponds to the liquid + under-saturated vapor state of the system.
However, it is not generally true that the liquid phase of nuclear matter
is characterized by a monotonic increase of the average isospin as function of fragment
size since, when the system is in liquid + saturated vapor coexistence, the largest fragment
in each event which for sure belongs to the liquid part does not show the same behavior.
Thus, taking into account that in the heavy fragment range of the mass spectrum
(where $<N/Z>$ decreases with $Z$)
the dominant fragment is the largest one in each event, is easy to anticipate
the fall of the average isospin of the largest cluster with its size.
This means that,
at least within the presently used microcanonical multifragmentation model,
fragment partition at break-up does not obey a total symmetry energy minimization principle
but is the consequence of the interplay of all observables entering
the statistical weight of a configuration - the key quantity of the model.

\section{Effect of source isospin on $<N/Z>$ vs. $Z$ distributions}

\begin{figure}
\resizebox{0.6\textwidth}{!}{%
  \includegraphics{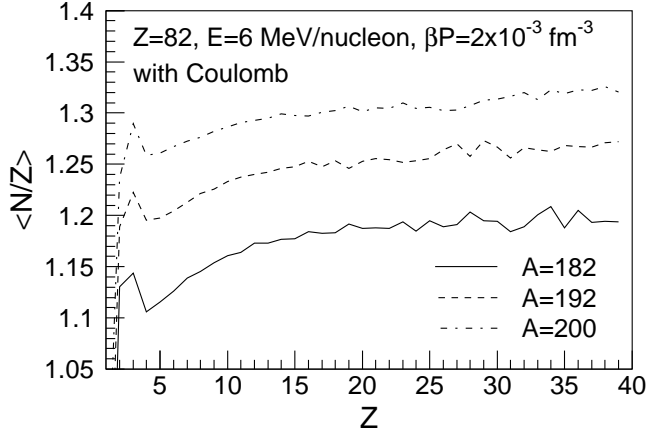}}
\caption{Fragment average isospin distributions corresponding to three sources
characterized by $Z$=82 and $A$=182, 192 and 200 with the
excitation energy 6 MeV/nucleon and $\beta P=2 \cdot 10^{-3}$ fm$^{-3}$. 
For all cases Coulomb and hard-core interactions are considered.}
\label{fig:iso_ef}
\end{figure}

To get a more complete picture on how sensitive the average isospin distributions
are with respect to the isospin of the source nucleus, $<N/Z>$ versus $Z$ distributions
have been analyzed in different regions of the phase diagram.
Because the results are similar irrespective the position of the
multifragmentation event inside the phase diagram, we restrict ourselves
to present only the most physically relevant case of a source with $Z$=82 
with Coulomb whose mass number was varied from 182 to 200. 
Here, the (i) freeze-out scenario was used.
Thus, Fig. \ref{fig:iso_ef} presents $<N/Z>$ versus $Z$ distributions
obtained for the excitation energy 6 MeV/nucleon and
$\beta P=2 \cdot 10^{-3}$ fm$^{-3}$. 
The result is intuitive: by increasing the source isospin, more neutrons are emitted
at break-up and all produced fragments are more neutron rich, leading to a shift of
the $<N/Z>$ versus $Z$ distributions toward higher values of $N/Z$
without any noticeable modification of their shape.

\section{Effect of secondary particle evaporation}

Even if all thermodynamical relevant information corresponds to the break-up stage of the
reaction, in the following we shall analyze 
the effect of secondary particle emission from primary excited fragments.
The motivation is that this reaction stage is the one accessible in
multifragmentation experiments. Sequential particle emission is treated using the
standard Weisskopf evaporation scheme as described in Ref. \cite{mmm}.

\begin{figure}
\resizebox{0.99\textwidth}{!}{%
  \includegraphics{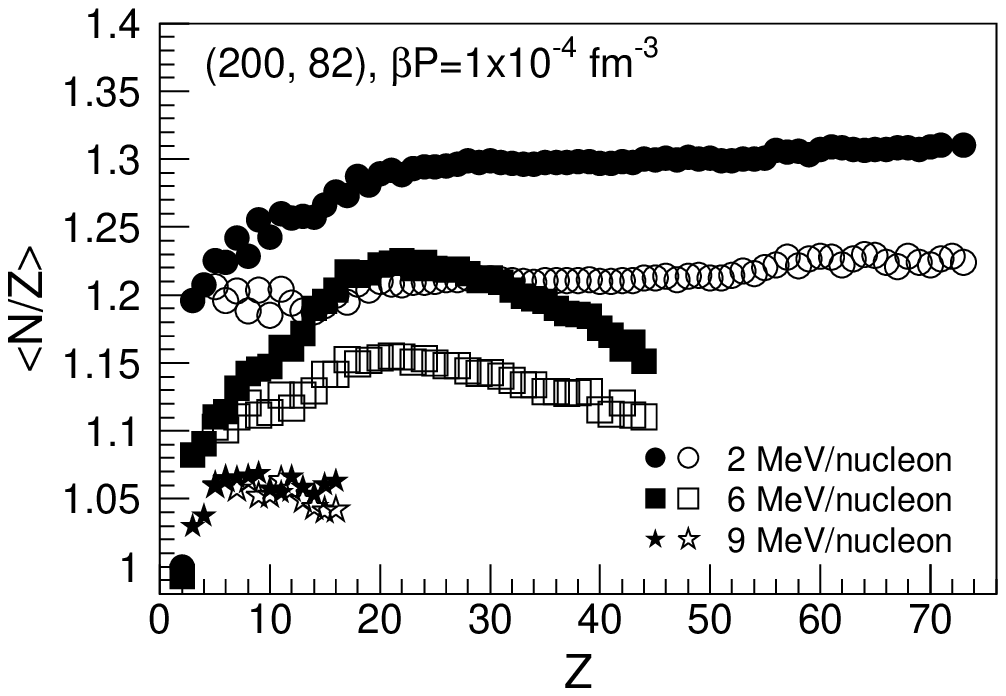}
  \includegraphics{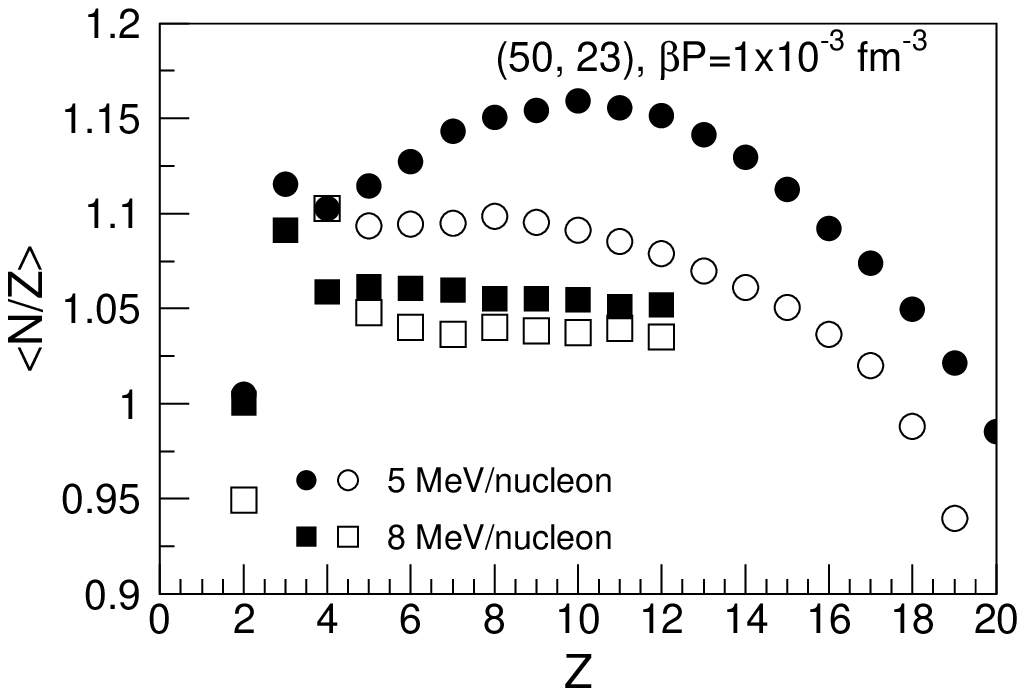}
  }
\caption{Fragment average isospin distributions in the break-up (close symbols)
and asymptotic (open symbols) stages of the reaction. Left panel corresponds to the case of 
(200, 82) nucleus, $\beta P=1 \cdot 10^{-4}$ fm$^{-3}$ with $E$=2, 6 and 9 MeV/nucleon;
Right panel corresponds to the case of 
(50, 23) nucleus, $\beta P=1 \cdot 10^{-3}$ fm$^{-3}$ with $E$=5,8 MeV/nucleon.
In all cases Coulomb interaction is present. The (ii) freeze-out scenario was used.}
\label{fig:evap_ef}
\end{figure}

Given that during the evaporation stage neutrons are emitted with the largest probability,
one may expect a more symmetric matter in the asymptotic stage of the reaction with respect
to the break-up stage. 
However, the dependence of evaporation probabilities
on both fragment average excitation energy and isospin makes 
the evaporation effect difficult to anticipate quantitatively
and raises the question whether or not isospin distributions
observed in the break-up stage survive in the asymptotic stage of the reaction.
To answer this question Fig. \ref{fig:evap_ef} 
illustrates the results obtained in different points of the phase diagram
in the physical case in which Coulomb interaction is present. 
The liquid + under-saturated vapor phase is represented by (200, 82) with
$\beta P=1 \cdot 10^{-4}$ fm$^{-3}$ and $E$=2 MeV/nucleon;
the liquid + saturated vapor coexistence is represented by 
(200, 82), $\beta P=1 \cdot 10^{-4}$ fm$^{-3}$ and $E$=6 MeV/nucleon and
(50, 23), $\beta P=1 \cdot 10^{-3}$ fm$^{-3}$, $E$=5 MeV/nucleon;
the super-saturated vapor phase is represented by
(200, 82) with $\beta P=1 \cdot 10^{-4}$ fm$^{-3}$, $E$=9 MeV/nucleon and 
(50, 23), $\beta P=1 \cdot 10^{-3}$ fm$^{-3}$, $E$=8 MeV/nucleon.
The conclusions are that for the liquid phase particle evaporation induces a lowering
of the fragment average isospin without modifying the linear dependence of
$<N/Z>$ on $Z$.
For the phase coexistence region, sequential evaporations act such that the rise and fall
shape of $<N/Z>$ versus $Z$ is diminished without being washed out. This effect is easy to
understand having in mind that neutron emission is stronger for the neutron-rich fragments
(maximum of $<N/Z>$ versus $Z$) with respect to the neutron-poor fragments from the extremities
of $<N/Z>$ vs. $Z$ distribution. 
The break-up fragments forming the super-saturated vapor phase
of are almost symmetric such that
sequential particle emission do not change to a significant extent their isotopic composition.
The asymptotic stage average isospin values corresponding to light charged particles ($Z<4$)
have been omitted from Fig. \ref{fig:evap_ef} because present some irregularities
interpreted as artefacts of the simplified procedure in which secondary decays have been
implemented \cite{mmm}.

\section{Do we have isospin fractionation in statistical multifragmentation models?}

Since isospin fractionation was decided by both dynamical models and
experimental multifragmentation data analyzing exclusively the isotopic content of
light charged emitted clusters,
it would be interesting to check to what extend our predictions agree
with the reported results.
To make this comparison straightforward we adopt one of the methods applied
by dynamical models and classify the fragments as part of liquid and gas phases
according to their mass.
Thus, in the spirit of Ref. \cite{ono}
we assume that the collection of fragments with $Z \leq 4$ form the "free" phase
and the rest of fragments form the "bound" phase. We prefer to call the obtained
subsystems by "free" instead of "gas" and "bound" instead of "liquid" because the
above mentioned classification may be done irrespectively the localization of the
multifragmentation event in the phase diagram, even outside the phase coexistence region. 
The choice of $Z=4$ as criterion for phase definition is arbitrary and similar results
have been obtained when $Z=2$ has been used.

\begin{figure}
\resizebox{0.99\textwidth}{!}{%
  \includegraphics{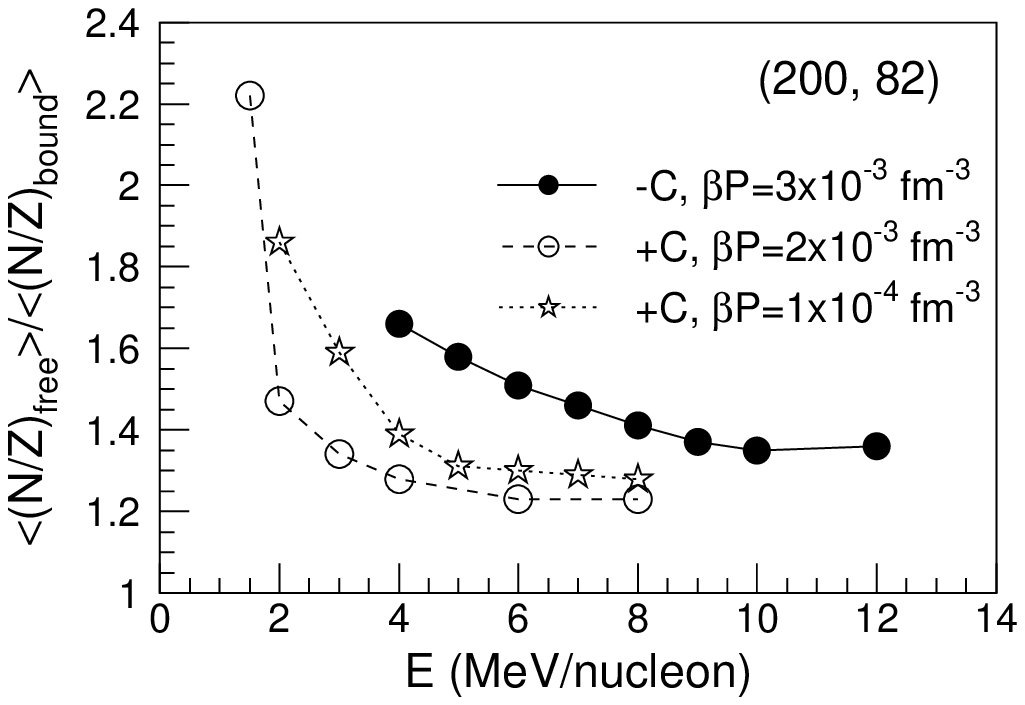}
  \includegraphics{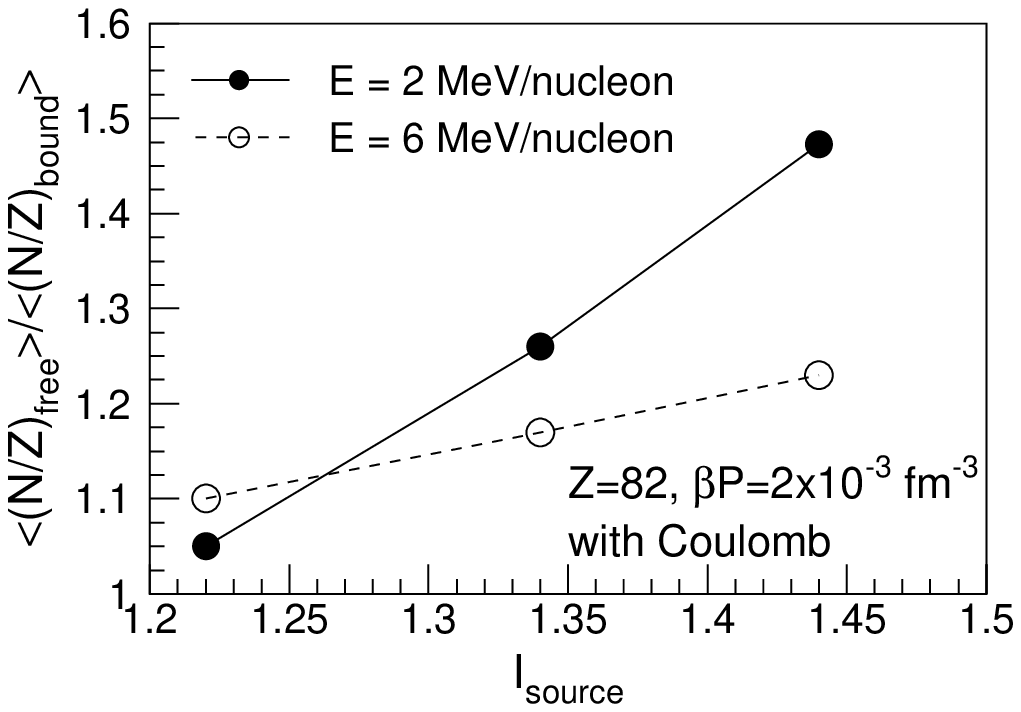}
  }
\caption{Left panel: Relative neutron enrichment of the "free" phase with respect
to the "bound" phase for the (200,82) nuclear system with and without Coulomb interaction
when the associated phase spaces are explored following the $\beta P$ constant paths
$2 \cdot 10^{-3}$ fm$^{-3}$, $1 \cdot 10^{-4}$ fm$^{-3}$
and respectively $3 \cdot 10^{-3}$ fm$^{-3}$;
Right panel: Relative neutron enrichment of the "free" phase with respect to the
"bound" phase as a function of source's isospin in the case of a system with $Z$=82, with
Coulomb interaction at 2 and 6 MeV/nucleon excitation energy and
$\beta P=2 \cdot 10^{-3}$ fm$^{-3}$.}
\label{fig:iso_f_b}
\end{figure}

The left panel of Fig. \ref{fig:iso_f_b}
depicts the relative neutron enrichment of the "free" phase with respect
to the "bound" phase for the (200, 82) nuclear system with (without) Coulomb
interaction along the $\beta P=2 \cdot 10^{-3}$ fm$^{-3}$ and $ 1 \cdot 10^{-4}$ fm$^{-3}$
($3 \cdot 10^{-3}$ fm$^{-3}$)
constant paths. As one may see, for all considered cases
$<N/Z>_{free}/<N/Z>_{bound}$ is larger than one, meaning that
isospin distillation takes place and its exact values depends on the observables
characterizing the state of the source. 
We mention at this point that relative neutron enrichment of the free phase
decreasing with increasing energy
was evidenced also by dynamical models \cite{baoanli},
a quantitative comparison being nevertheless
impossible because of different definitions of phases.

The right panel of Fig.\ref{fig:iso_f_b}
represents the dependence of the relative enrichment of the "free" phase as a function of
source isospin.
As expected, more isospin asymmetric is the initial system, more neutron rich is the
corresponding "free" phase obtained by multifragmentation.
Similar linear behavior was reported by dynamical models \cite{baoanli}.

An important remark to be made is that, at least in the framework of the
presently used microcanonical multifragmentation model,
the neutron enrichment of the ``free'' phase with respect to the 
``bound'' phase is not necessarily connected with phase coexistence since manifests
even for the (200, 82), with Coulomb interaction case where,
for freeze-out volumes smaller than several tens $V_0$,
the employed model does not
enter phase coexistence region (left panel of Fig. \ref{fig:iso_f_b}).

\section{Conclusions} 

Fragment isospin distributions have been investigated in the framework of a
microcanonical multifragmentation model.
The obtained distributions manifest different shapes in the liquid, liquid-gas and gas regions of
the phase diagram.
For small systems with low excitation energies mass and charge conservation constraints modify
the shape of $<N/Z>$ versus $Z$
distributions from a monotonically increasing one to a one having a maximum.
Distributions of the largest fragment in each event expected to be a good order parameter of the
phase transition and a reliable estimation of the liquid part does not manifest bimodality in the
whole phase coexistence zone and confirm the conclusions of Ref. \cite{gulminelli} stating that
mass conservation induces modifications of the $Y(Z_{max})$ distributions from the one obtained in
infinite systems.
Moreover, for low excitation energies $Y(Z_{max})$ shows bimodality outside the
coexistence region due to the interplay between evaporation and fission as decay
mechanisms.
Isospin fractionation is found to be compatible with a microcanonical multifragmentation
scenario and effects of secondary particle emission and isospin of the source are investigated.
Phase classification according to the cluster size proves that
neutron enrichment of the ``free'' phase with respect to the ``bound'' one depends
on the source state.  
Finally, neutron enrichment of the "free" phase is not a signal of
phase coexistence since it is observed everywhere in the phase space.


\begin{thebibliography}{99}

\bibitem{muller} H. Muller and B. D. Serot, Phys. Rev. C {\bf 52}, 2072 (1995).
\bibitem{baoanli} Bao-An Li, Phys. Rev. Lett. {\bf 85}, 4221 (2000).
\bibitem{sil} Tapas Sil, S. K. Samaddar, J. N. De and S. Shlomo,
              Phys. Rev. C {\bf 69}, 014602 (2004).
\bibitem{lee} S. J. Lee and A. Z. Mekjian, Phys. Rev. C {\bf 68}, 014608 (2003).
\bibitem{ono} Akira Ono, P. Danielewicz, W. A. Friedman, W. G. Lynch and M. B. Tsang,
              Phys. Rev. C {\bf 68}, 051601(R) (2003).
\bibitem{ditoro} V. Baran, M. Colonna, M. Di Toro, V. Greco, M. Zielinska-Pfabe and
                    H. H. Wolter, Nucl. Phys. {\bf A703}, 603 (2002).
\bibitem{xu} H. S. Xu {\it et al.}, Phys. Rev. Lett. {\bf 85}, 716 (2000).
\bibitem{geraci} E. Geraci {\it et al.}, Nucl. Phys. {\bf A732}, 173 (2004).
\bibitem{martin} E. Martin {\it et al.}, Phys. Rev. C {\bf 62}, 027601 (2000).
\bibitem{shetty} D. V. Shetty {\it et al.}, Phys. Rev. C {\bf 68}, 021602(R) (2003).
\bibitem{mmm} Al. H. Raduta and Ad. R. Raduta, Phys. Rev. C {\bf 55}, 1344 (1997); 
              {\it ibid.} Phys. Rev. C {\bf 65}, 054610 (2002).
\bibitem{gross} D. H. E. Gross and M. E. Madjet, cond-mat/9611192.
\bibitem{mmmc} D. H. E. Gross, Rep. Progr. Phys. {\bf 53}, 605 (1990).
\bibitem{smm} J. P. Bondorf, A. S. Botvina, A. S. Iljinov, I. N. Mishustin and K. Sneppen,
                  Phys. Rep. {\bf 257}, 133 (1995).
\bibitem{randrup} S. E. Koonin and J. Randrup, Nucl. Phys. A {\bf 474}, 173 (1987).
\bibitem{gulminelli_const_bp} F. Gulminelli, Ph. Chomaz and V. Duflot,
                                    Europhys. Lett. {\bf 50}, 434 (2000).
\bibitem{parlog} M. Parlog {\it et al.}, Eur. Phys. J. A {\bf 25}, 223 (2005).
\bibitem{prl2001} Al. H. Raduta and Ad. R. Raduta, Phys. Rev. Lett. {\bf 87}, 202701 (2001).
\bibitem{prl2003} F. Gulminelli, Ph. Chomaz, Al. H. Raduta and Ad. R. Raduta, 
                  Phys. Rev. Lett. {\bf 91}, 202701 (2003).
\bibitem{campi} X. Campi, J. Desbois and E. Lipparini, Phys. Lett. {\bf B138}, 353 (1984).
\bibitem{purdue}J. B. Elliott {\it et al.}, Phys. Rev. C {\bf 62}, 064603 (2000).
\bibitem{gulminelli} F. Gulminelli and Ph. Chomaz, Phys. Rev. C {\bf 71}, 054607 (2005).
\bibitem{ld}  W. D. Myers and W. J. Swiatecki, Nucl. Phys. {\bf 81}, 1 (1967);
{\it ibid.} Ark. Fiz. {\bf 36}, 343 (1967).
\bibitem{shetty_botvina} D. V. Shetty, A. S. Botvina, S. J. Yennello, A. Keksis, E. Martin and 
                         G. A. Souliotis, nucl-ex/0401012.  
\bibitem{texas} G. A. Souliotis, D. V. Shetty, M. Veselsky, G. Chubarian, L. Trache, A. Keksis,
                E. Martin and S. J. Yennello, Phys. Rev. C {\bf 68}, 024605 (2003).


\end{thebibliography}
\end{document}